\documentclass{appolb}
\usepackage{graphicx}
\usepackage{cite}
\usepackage{hyperref}
\usepackage{amsmath}
\usepackage{epstopdf}%
\usepackage{amsfonts}%
\usepackage{amssymb}
\usepackage{tikz}
\usepackage{booktabs}
\usepackage{wrapfig}
\DeclareMathOperator{\disc}{Disc}

\begin{document}

% \eqsec  % uncomment this line to get equations numbered by (sec.num)
\title{Dynamical generation of hadronic resonances%
\thanks{Presented at the Workshop ``EEF70 -- Workshop on Unquenched Hadron Spectroscopy: Non-Perturbative Models and Methods of QCD vs. Experiment'', Coimbra, Portugal, September 1--5, 2014.}%
% you can use '\\' to break lines
}
\author{T. Wolkanowski
\address{Institut f\"{u}r Theoretische Physik, Goethe-Universit\"at Frankfurt am Main Max-von-Laue-Str. 1, 60438 Frankfurt am Main, Germany}
}
\maketitle

\begin{abstract}
One type of dynamical generation consists in the formation of multiple hadronic resonances from single seed states by incorporating hadronic loop contributions on the level of $s$-wave propagators. Along this line, we study the propagator poles within two models of scalar resonances and report on the status of our work: (i) Using a simple quantum field theory describing the decay of $f_{0}(500)$ into two pions, we may obtain a second, additional pole on the first Riemann sheet below the pion-pion threshold (\emph{i.e.}, a stable state can emerge). (ii) We perform a numerical study of the pole(s) of $a_{0}(1450)$ by using as an input the results obtained in the extended Linear Sigma Model (eLSM). Here, we do not find any additional pole besides the original one, thus we cannot obtain $a_{0}(980)$ as an emerging state. (iii) We finally demonstrate that, although the coupling constants in typical effective models might be large, the next-to-leading-order contribution to the decay amplitude is usually small and can be neglected.
\end{abstract}

  
\section{Introduction}
From various experimental and theoretical studies during the past three decades it became clear that the scalar hadronic states cannot be incorporated into the ordinary $q\bar{q}$ picture based on a simple representation of $SU(3)$ flavour symmetry \cite{amslerrev,beringer}. Some related questions are: Why are $f_{0}(500)$ (or $\sigma$) and $K_{0}^{\ast}(800)$ (or $\kappa$) so broad? Why are $f_{0}(980)$ and $a_{0}(980)$ so much narrower? Why does it seem that we find much more scalar--isoscalar states than we expected ($f_{0}(500)$, $f_{0}(980)$, $f_{0}(1370)$, $f_{0}(1500)$ and $f_{0}(1710)$)? Dealing with these problems, models of \emph{dynamical generation}\footnote{For a general discussion concerning dynamical generation see Ref. \cite{giacosaDynamical} and references therein.} began to offer one possible solution and hence received more and more attention from the community \cite{dullemond,2006beveren,tornqvist,boglione,pennington} (for other approaches, see \emph{e.g.} Refs. \cite{jaffe,maiani,hooft,fariborz,rodriguez,isgur,tqmix,oller}). Models of dynamical generation focus on the unitarization of bare scalar (seed) states via strong couplings to intermediate (hadronic) states. (One sometimes also speaks in this context of mesonic dressing or the influence of mesonic clouds \cite{salam,achasov,giacosaSpectral}.) According to them, because of the coupling to hadronic channels, the scalar sector not only escapes from the general approach of the naive quark model, but additional resonances with the same quantum numbers can be generated as poles on the unphysical Riemann sheet, usually obtained in the scattering matrix.

In contrast, the mass and decay width of a resonance may be determined by the pole of its full interacting propagator, too -- a procedure first proposed by Peierls \cite{peierls} long ago (see also Refs. \cite{levypoles,aramaki,landshoff}). The idea of dynamical generation then means to look -- besides for a pole coming from the seed sate -- for companion poles on Riemann sheets which are accessable due to the allowed decay channels. For instance, a reasonable model giving the propagator of the scalar--isovector state $a_{0}(1450)$ could automatically yield the corresponding narrow state below $1$ GeV, the $a_{0}(980)$ \cite{dullemond,2006beveren,tornqvist,boglione,pennington}.

Following this line, we calculate in this paper the propagators of scalar resonances within two effective models and study the mechanism of dynamical generation by analysing the poles on the appropriate Riemann sheet(s). Then, to make a connection to the presented models, we summarize a recent work which emphasizes that the next-to-leading order (NLO) contribution to the decay amplitude is usually negligible in the case of effective hadronic models.

\section{Used method}
The one-loop approximation yields the inverse propagator of a scalar resonance after applying Dyson resummation of the hadronic loop contributions:
\begin{equation}
\Delta^{-1}(s)=s-M_{0}^{2}-\Pi(s) \ ,
\end{equation}
where $\Pi(s)=\sum_{i}\Pi_{i}(s)$ is the sum of all included channels and $M_{0}$ is the bare or tree-level mass. The propagator on the unphysical sheet(s) is obtained by analytic continuation according to
\begin{equation}
\Pi_{i}^{c}(z) = \Pi_{i}(z)+\disc\Pi_{i}(z) \ , \ \ \disc\Pi_{i}(s) = 2i\lim_{\epsilon \to 0^{+}}\operatorname{Im}\Pi_{i}(s+i\epsilon) \ .
\end{equation}
Here, the superscript $c$ indicates the continued function on the next sheet. The decay amplitude for each channel is calculated from the relevant interaction terms of the underlying Lagrangian. The optical theorem for Feynman diagrams can then be applied to compute the imaginary part of the corresponding self-energy loop $\Pi_{i}(s)$, regularized by a Gaussian 3d-cutoff function with some cutoff scale $\Lambda$ (since we deal with effective models in the low-energy regime to study light mesons, it is reasonable to set $\Lambda$ between 1 and 2 GeV). The real part is finally obtained by the dispersion relation
\begin{equation}
\operatorname{Re}\Pi_{i}(s) = \frac{1}{\pi} -\!\!\!\!\!\!\!\int\text{d}s^{\prime} \ \frac{-\operatorname{Im}\Pi_{i}(s^{\prime})}{s-s^{\prime}} \ .
\end{equation}
Note that in this paper the appropriate sheet to look for poles is taken to be the one closest to the physical region.

\section{Results and discussions}
\subsection{Simple toy model for $f_{0}(500)$}
Our starting point is a toy model with two scalar fields, $S$ and $\phi$, representing the $f_{0}(500)$ state and (neutral) pions, and containing the one-channel decay process $S\rightarrow\phi\phi$:
\begin{equation}
\mathcal{L}=\frac{1}{2}(\partial_{\mu}S)^{2}+\frac{1}{2}(\partial_{\mu}\phi)^{2}-\frac{1}{2}M_{0}^{2}S^{2}-\frac{1}{2}m^{2}\phi^{2}+gS\phi^{2} \ .
\label{eq:SphiphiLagrangian}
\end{equation}
For general studies of this model see Refs. \cite{veltman,giacosaSpectral}, while detailed works concerning its pole structure can be found in Refs. \cite{e38,thomasthesis}.

By fixing $\Lambda$, the remaining free parameters $M_{0}$ and the coupling constant $g$ can be determined if one takes the well-known $\sigma$-pole of Caprini \emph{et al.} \cite{caprini} as a correct determination of the $\sigma$-mass and -width. This then requires an additional pole slightly below the two-pion threshold, corresponding to a stable state which is dynamically generated by hadronic interactions. Its trajectory is quite interesting: it is first situated on the real axis of the second sheet for small and intermediate values of $g$, then vanishes for a coupling large enough by slipping into the branch cut and finally appears on the real axis of the first sheet, see Tab. \ref{tab:tab1}. Note that no experiment has yet indicated a hadronic particle with a mass lower than the two-pion threshold. We thus cannot exclude the possibility of having found a spurious pole (see also Ref. \cite{e38} and references therein).
\begin{table}[t]
\center
\begin{tabular}
[c]{lcccc}
\toprule & \multicolumn{2}{c}{$\Lambda=1$ GeV} & \multicolumn{2}{c}{$\Lambda=2$ GeV}\\
\midrule Pole $\sqrt{s}=M-i\hspace{0.02cm}\Gamma/2$ [MeV] & $M_{0}$ [GeV] & $g$ [GeV] & $M_{0}$ [GeV] & $g$ [GeV]\\
\midrule &  &  &  & \\
Sheet II: \ $441-i\hspace{0.02cm}272$ & 0.416 & 1.574 & 0.489 & 1.584\\
&  &  &  & \\
\midrule Sheet I: & \multicolumn{2}{c}{$0.257-i\hspace{0.02cm}\epsilon$} & \multicolumn{2}{c}{$0.269-i\hspace{0.02cm}\epsilon$}\\
&  &  &  & \\
\end{tabular}
\caption{Numerical results for the bare mass parameter $M_{0}$ and the coupling constant $g$ in dependence of the cutoff $\Lambda$. The additional pole below threshold is situated on the real axis of the first sheet.}
\label{tab:tab1}
\end{table}

\subsection{Scalar--isovector sector in the extended Linear Sigma Model}
The extended Linear Sigma Model (eLSM) is an effective model of QCD with (pseudo)scalar as well as (axial-)vector states based on chiral $U(3)_{L}\times U(3)_{R}$ symmetry and dilatation invariance \cite{eLSM2010(1),eLSM1,eLSM2}. Explicit (due to non-vanishing quark masses) as well as spontaneous symmetry breaking (due to a non-vanishing chiral condensate $\langle q\bar{q}\rangle\neq0$) and the $U(1)_{A}$ chiral anomaly are taken into account. In this model, the scalar isotriplet was identified as the resonance $a_{0}(1450)$ with a fitted tree-level mass $M_{0}=1363$ MeV. The relevant interaction part of the Lagrangian for the neutral state $a_{0}^{0}$ reads
\begin{equation}
\mathcal{L}_{\text{int}} = Aa_{0}^{0}\eta\pi^{0}+Ba_{0}^{0}\eta^{\prime}\pi^{0}+Ca_{0}^{0}(K^{0}\bar{K}^{0}-K^{-}K^{+}) \ ,
\end{equation}
where $\pi^{0},\eta,\eta^{\prime},K$ are the pseudoscalar mesons and the constants $A,B,C$ are combinations of the coupling constants and masses taken from Ref. \cite{eLSM2} (values on-shell).

For any reasonable cutoff parameter $\Lambda$ it turns out that the complex propagator pole on the sheet nearest to the physical region is too close to the real axis, hence yields a decay width that is too small. This is not in agreement with neither the tree-level result from the model, nor the experiment. The inclusion of loops in the way as we described above spoils the tree-level result (at least for the decay width). One should perform a reanalysis with full $s$-dependence. 

However, we do \emph{not} find a companion pole of $a_{0}(1450)$. This result does not change upon variations of the parameters. This may indicate that one should try to include the $a_{0}(980)$ as a tetraquark state into the eLSM and/or perform a full scattering analysis.

\subsection{The role of the next-to-leading order triangle-shaped diagram in two-body hadronic decays}
\begin{wrapfigure}{r}{0.3\textwidth}
\vspace{-0.25cm}
\begin{center}
\includegraphics[scale=0.55]{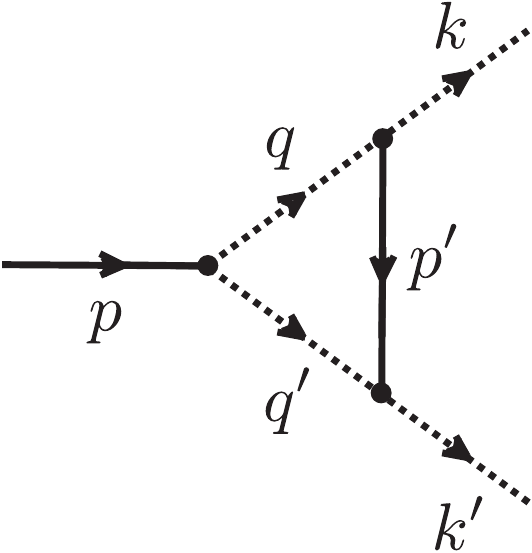}
\end{center}
\caption{Triangle-shaped NLO diagram for a two-body decay.}
\label{fig:fig1}
\end{wrapfigure}

When dealing with effective hadronic models, the question about the role of the next-to-leading order (NLO) diagram for the decays becomes relevant. Is it reasonable to discard this term, for example, as done in the model fit of the eLSM in the case of $a_{0}(1450)$? The diagram (see Fig. \ref{fig:fig1}) is proportional to the third power of the coupling constant, which in effective models is usually not a small number.

We evaluate the role of the triangle diagram for the simple scalar field theory of Eq. (\ref{eq:SphiphiLagrangian}) and then adopt the general result to the $\pi\pi$-decays of $f_{0}(500)$, $f_{0}(980)$, $f_{0}(1370)$, $f_{0}(1500)$ and $f_{0}(1710)$ as well as the $\bar{K}K$-decays of $a_{0}(1450)$ and $f_{0}(1710)$, for details see Ref. \cite{JonasNucl}. It turns out that, with the exception of $f_{0}(500)$, the NLO correction is negligible.

\bigskip

\textbf{Acknowledgments:} The author thanks J. Schneitzer and F. Giacosa for cooperation, and D. H. Rischke and J. Wambach for useful discussions. Financial support from HGS-HIRe, F\&E GSI/GU and HIC for FAIR Frankfurt is acknowledged.

\end{document}